\documentclass[preprint,a4paper,showkeys,nofootinbib]{revtex4}
\usepackage{graphicx}

\begin{document}

\newcommand{\be}{\begin{equation}}
\newcommand{\ee}{\end{equation}}
\newcommand{\bq}{\begin{eqnarray}}
\newcommand{\eq}{\end{eqnarray}}
\newcommand{\bc}{\begin{center}}
\newcommand{\ec}{\end{center}}
\newcommand{\pd}{\partial}
\newcommand{\Mpl}{{\mathrm{Mpl}}}

\title{The effect of kinematic constraints on multi-tension string network evolution}

\author{Anastasios Avgoustidis\footnote{E-mail: a.avgoustidis@damtp.cam.ac.uk}$^{a,b}$, Edmund.~J. Copeland\footnote{E-mail: ed.copeland@nottingham.ac.uk}$^{c}$}
\affiliation{$^a$ Institute of Cosmos Sciences, Faculty of Physics, University of Barcelona, Marti i Franques 1, 08024 Barcelona, Spain\\ $^b$ Centre for Theoretical Cosmology, DAMTP, CMS, Wilberforce Road, Cambridge CB3 0WA, UK\\ 
$^c$School of Physics and Astronomy, University of Nottingham, University Park, Nottingham NG7 2RD, UK}

\date{\today}

\begin{abstract}
We consider the evolution of a network of strings in an expanding universe, allowing for the formation of junctions between strings of different tensions. By explicitly including, in the velocity-dependent evolution equations for the network, kinematic constraints associated with the formation of Y-shaped string junctions, we show how they lead to scaling solutions in regimes where they would not otherwise be found, thereby extending the range of parameters which lead to scaling. By incorporating these constraints we are able to study their general behaviour for networks with cosmic superstring interaction rules, and predict the scaling densities expected by these networks.  
\end{abstract}


\maketitle

\section{Introduction}
Understanding how a network of cosmic strings evolves in our universe has been an area of interest for a number of decades following the pioneering work of Kibble  \cite{Kibble:1980mv,Kibble} (for a review  see \cite{book,Hindmarsh:1994re}). Being a complicated non-linear system, various approaches have been adopted including numerical simulations \cite{Bennett:1989yp,Allen:1990tv} as well as analytic approximations \cite{Kibble,AusCopKib,VOS,VOSk}. The interest in the results of these simulations wained when it became evident that cosmic strings by themselves were never going to provide the seed primordial density fluctuations~\cite{Albrecht:1997nt}, but recently interest in them has resurfaced thanks to the realisation that in a reasonably large class of Superstring motivated models it is possible to form a network of strings \cite{JST,Jones:2002cv,Sarangi:2002yt,Majumdar:2002hy}, that can live as long as the age of the Universe, be of cosmological length and survive a period of early universe inflation.  Moreover these cosmic superstrings could possibly be detected, thereby providing the first direct evidence of string theory in nature (for recent reviews of cosmic superstrings see  \cite{Polchinski:2004ia,Davis:2005dd,Sakellariadou:2009ev,Myers2009,Copeland:2009ga}). This has spawned a renewed interest in the subject, in particular there is now a need to understand how a network of cosmic superstrings evolves. The fact that there can now be more than one type of string has added a new ingredient, the possibility of forming junctions when two different types intersect, the junction marking the point where a third string emerges. A number of approaches have already been adopted to study this problem. On the numerical side, network evolution with junctions has been studied in~\cite{McGraw,SpergPen,CopSaf,Urrestilla:2007yw}.  Preliminary work based on non-abelian field theory models had suggested that networks with junctions may be obstructed from scaling~\cite{SpergPen} (network frustration), but, more recently, analytic~\cite{Martins_nonint,TWW,NAVOS} and numerical~\cite{CopSaf,HindSaf,Urrestilla:2007yw} evidence has accumulated favouring a picture in which these networks\footnote{Similar results are also found in field theory simulations of domain wall networks with junctions~\cite{Avelino1,Avelino2,Avelino3}.}
reach scaling, while the various network components are allowed to interact with each other forming 3-string Y-type junctions. The development of these field-theory analogues of cosmic superstrings have involved either combinations of interacting abelian models \cite{Saffin:2005cs,Rajantie:2007hp,Urrestilla:2007yw}  or through non-abelian models \cite{SpergPen,McGraw,CopSaf,HindSaf}. In both cases the networks that form admit trilinear vertices, hence junctions where the usual intercommutation properties of the strings no longer apply.

On the analytic front, the first successful, comprehensive model was that of~\cite{TWW}, which was based on the velocity-dependent string evolution models of Martins and Shellard \cite{VOS,VOSk}. In~\cite{TWW} the authors assign a single correlation length and velocity to all network components and follow the evolution of the number density of each of the different species, finding generic scaling under the assumption that the energy associated to the formation of junctions is rapidly radiated away and decouples from the string network.  In order to make the basic model more realistic, in reference~\cite{NAVOS}, the authors extended the velocity-dependent models by associating a different correlation length and velocity to each network, and enforcing energy conservation at junction formation.   
While once again, scaling solutions are readily found in this model, and are in qualitative agreement with those of reference~\cite{TWW}, the scaling behaviour is seen not to be generic in the model of \cite{NAVOS}. In fact, for scaling to occur, it requires that the terms describing the formation of junctions be subdominant -- or at most comparable in magnitude -- to the corresponding self-interaction terms. 

As well as understanding the dynamics of the string network, properly incorporating the kinematics of strings that can form junctions is equally important and this area has recently started receiving attention~\cite{CopKibSteer1,CopKibSteer2,Davis:2008kg}. Studies of the collisions of Nambu-Goto strings with junctions at which three strings meet have shown that the exchange to form junctions cannot occur if the strings meet with very large relative velocity. For the case of non-abelian strings, rather than passing through one another they become stuck in an X configuration 
\cite{CopKibSteer2,Carter:2006cf}, in each case the constraint depending on the angle at which the strings meet, on their relative velocity, and on the ratios of the string tensions. Calculations of the average speed at which a junction moves along each of the three strings from which it is formed yield results consistent with the observation that junction dynamics may be such as to preferentially remove the heavy strings from the network leaving a network of predominantly light strings.  In \cite{CopFirKibSteer} the authors extended the analysis to include the formation of three-string junctions between $(p,q)$-cosmic superstrings, which required modifications of the Nambu-Goto equations to take account of the additional requirements of flux conservation. Investigating the collisions between such strings they showed that kinematic constraints analogous to those found previously for collisions of Nambu-Goto strings apply here too. 

A number of these  constraints that have emerged from analysing the modified Nambu-Goto equations have been checked through numerical simulations of field-theory strings which can also lead to junction formation \cite{Sakellariadou:2008ay,Salmi,Bevis:2008hg,Bevis:2009az}. In most cases good agreement has been obtained with the analytical predictions, although the results of \cite{Urrestilla:2007yw} show some surprising features, and of course some important differences emerge due to the fact that field-theory strings have interactions between them that are not found in the Nambu-Goto case, hence some of the detailed numbers differ. 

In this paper we show how to consistently include the kinematic constraints associated with Y-shaped string junctions~\cite{CopKibSteer1,CopKibSteer2,CopFirKibSteer} into the model of \cite{NAVOS}.   A first step towards this goal has been taken in reference~\cite{Cui:2007js} by introducing a velocity cut-off beyond which colliding strings fail to exchange partners.  Here, we take into account the dependence of the constraints on both the velocity and orientation of the colliding strings, and demonstrate how these constraints provide the required suppression factors for the relevant junction-formation terms, thereby facilitating scaling in regimes where scaling behaviour would not otherwise be observed.  By incorporating these constraints in the velocity-dependent evolution models we are able to study their general behaviour for networks with cosmic superstring interaction rules. 

The rest of the paper is as follows. In section~\ref{models} we introduce the velocity-dependent string evolution models, and in section~\ref{constrs} we describe the kinematic constraints that apply to $(p,q)$-junction formation in cosmic superstrings. This is followed in section~\ref{incorp} with the incorporation of the constraints in the  string evolution models. Our key results are presented in section~\ref{results} before we conclude in section~\ref{discuss}.

\section{\label{models}Velocity-Dependent String Evolution Models}

Let us begin by briefly reviewing the model of reference~\cite{NAVOS}, which will be the starting point of our present discussion.  This is a phenomenological model for the quantitative description of the cosmological evolution of cosmic string networks that are composed of different string types, generally of unequal  tensions, allowed to interact among each other forming trilinear vertices (Y-type junctions).  Each network component $i$ is assumed to be Brownian (that is of a random walk structure) and is characterised by its tension $\mu_i$, correlation length $L_i$ and root-mean-squared (RMS) velocity $v_i$, the latter two quantities being functions of cosmological time $t$.  Due to the Brownian structure of the networks, the correlation length $L_i(t)$ defines, at any time $t$, a network energy density component $\rho_i(t)=\mu_i/L^2_i(t)$.  

Starting from the Nambu-Goto action in a Friedmann-Lema{\^i}tre-Robertson-Walker (FLRW) background with scale factor $a(t)$ and averaging over the string worldsheet, one can derive evolution equations for $\rho_i$ and $v_i$ that account for string stretching and dilution as well as velocity redshifting due to cosmic expansion.  One then introduces additional phenomenological terms~\cite{Kibble,VOSk,NAVOS} describing the interaction among strings of the same type, producing loops, as well as cross interactions among different types of strings leading to the formation of Y-type junctions with new segments (links/zippers) connecting the colliding strings. The resulting equations are~\cite{NAVOS}:              

\be\label{rho_idt}
  \dot\rho_i = -2\frac{\dot a}{a}(1+v_i^2)\rho_i-\frac{\tilde c_i
  v_i\rho_i}{L_i} - \sum_{a,k} \frac{\tilde d_{ia}^k
  \bar v_{ia} \mu_i \ell_{ia}^k(t)}{L_a^2 L_i^2} + \sum_{b,\,a\le b}
  \frac{\tilde d_{ab}^i \bar v_{ab} \mu_i
  \ell_{ab}^i(t)}{L_a^2 L_b^2}
\ee
\be
  \dot v_i = (1-v_i^2)\left[\frac{k_i}{R_i}-2\frac{\dot a}{a}v_i
  + \sum_{b,\,a\le b} \tilde b_{ab}^i \frac{\bar v_{ab}}{v_i}
  \frac{(\mu_a+\mu_b-\mu_i)}{\mu_i}\frac{\ell_{ab}^i(t)
  L_i^2}{L_a^2 L_b^2}\right] \, , \label{v_idt}
\ee
where $\tilde c_i$ parametrises the self-interaction efficiency of strings of type $i$ producing loops\footnote{This can be expressed as an integral of an appropriate loop production function over all relevant loop sizes~\cite{book}.} and $\tilde d_{ij}^k=\tilde d_{ji}^k$ denotes the efficiency parameter for the process in which strings of type $i$ and $j$ interact to produce a segment of type $k$. The mean velocity $\bar v_{ij}$ is the average relative RMS velocity 
between strings of type $i$ and $j$, and $\ell_{ij}^k(t)$ is the average length of links/zippers of type $k$, produced by interactions between strings of types $i$ and $j$ around time $t$.  In the second equation, $k_i$ is the curvature parameter, which indirectly encodes information about the small-scale structure on strings and can be expressed as a function of the velocity $v_i(t)$.  In the relativistic limit one finds~\cite{VOSk}:
\be\label{k}
k_i=\frac{2\sqrt{2}}{\pi}\frac{1-8v_i^6}{1+8v_i^6}\,.
\ee  
Under the Brownian network assumption, the radii of curvature $R_i$ are equal to the corresponding correlation lengths $L_i$, so one should set $R_i=L_i$ in equation (\ref{v_idt}).  Finally, the parameters $\tilde b_{ij}^k$ are introduced here in order to be able to switch between the model of reference~\cite{NAVOS} (where $\tilde b_{ij}^k=\tilde d_{ij}^k$), in which the energy liberated by the formation of junctions is redistributed in the network as kinetic energy, and a model analogous to reference~\cite{TWW} (corresponding to $\tilde b_{ij}^k=0$), where all this energy is assumed to be radiated away.    

One can use this general class of models in different contexts ($Z_N$-strings, non-abelian strings, cosmic superstrings) to describe networks with multiple string components that interact with each other according to rules encoded in the coefficients $\tilde d_{ij}^k, \tilde c_i$.  The string tensions $\mu_i$, interaction rules $\tilde d_{ij}^k, \tilde c_i$ and link/zipper length $\ell_{ij}^k(t)$ are model dependent.     

In the case of cosmic superstrings, for example, the interaction rules are well-understood~\cite{JackJoPolch,TWW,Jackson}.  There are three possible outcomes of the crossing of a $(p,q)$ and a $(p',q')$ string: they can either pass through one other or zip in two different ways, producing either a $(p+p',q+q')$ or a $(p-p',q-q')$ zipper segment.  Given that the strings have zipped, the probability that the additive/subtractive channel is followed is given by:            
\be\label{Paddsubtr}
 P_{(p,q),(p',q')}^\pm = \frac{1}{2}\left( 1\mp\left( \frac{pp'g_s^2
 +qq'}{(p^2g_s^2+q^2)^{1/2}\,(p'^2g_s^2+q'^2)^{1/2}} \right) \right) ,
\ee
where we have assumed that the Ramond-Ramond scalar present in the analysis of \cite{JackJoPolch} is zero.  Therefore, the coefficients $\tilde d_{ij}^k$ in equation (\ref{rho_idt}) can be expressed for the case of cosmic superstrings as: 
\be\label{d_super}
 \tilde d_{(p,q),(p',q')}^{(p\pm p',q\pm q')} =
 \tilde d_{(p,q),(p',q')} P_{(p,q),(p',q')}^\pm ,
\ee
where the factor $\tilde d_{(p,q),(p',q')}$ corresponds to the probability that the crossing is non-trivial (i.e. the strings do not pass through one other but rather form a zipper following either of the two above mentioned channels).  This factor depends on the compactification and string coupling $g_s$, and for different string types lies in different ranges, which have been investigated in detail in references~\cite{JackJoPolch,Jackson}.  We will treat these factors as free parameters chosen in the appropriate ranges, which typically lie between $10^{-3} \leq \tilde d_{(p,q),(p',q')} \leq 1.$    

The string tensions are also known. For example, for a string carrying charges ($p_i\,,q_i$), in flat spacetime, its tension is given by:
\be\label{tensions}
\bar \mu_{i} =\frac{\mu_F}{g_s}\sqrt{p_i^2g_s^2+q_i^2}
\ee
where $\mu_F$ is the tension of the fundamental (1,0) string (F-string) and, again, we have assumed the Ramond-Ramond scalar to be zero.   

To close the system one also needs to specify a functional form for the average length of the produced links/zippers $\ell_{ij}^k(t)$.  For superstrings the formation of junctions happens by zipping along the length of colliding strings, so the natural choice is to take $\ell_{ij}^k(t)$ equal to the smallest of the two correlation lengths $L_i(t)$, $L_j(t)$.  As in~\cite{NAVOS}, we will use  
\be\label{ell}
\ell_{ij}^k=\frac{L_i L_j}{L_i+L_j} \, , 
\ee
a simple expression returning a value smaller than (but close to) the smallest of the two correlation lengths.  Other choices can be made, depending on the model, but as long as $\ell_{ij}^k(t)$ scales as some mean of the correlation lengths, these choices do not significantly alter the results~\cite{NAVOS}, especially in the case when the junction formation terms are subdominant, as will be the case here. 

Equations (\ref{rho_idt}-\ref{ell}), truncated at any finite number of string species, can be solved numerically for any choice of the free parameters $\tilde c_i$, $\tilde d_{ij}^k$.  We now proceed to describe the microphysical kinematic constraints for Y-junction formation, which we will incorporate into our string evolution models in section~\ref{incorp}.  As we will see, including these constraints in our models will lead to a systematic suppression of the junction formation parameters $\tilde d_{ij}^k$.

\section{\label{constrs}Kinematic Constraints for $(p,q)$-junction formation}

In \cite{CopFirKibSteer} the authors extended arguments originally presented in \cite{CopKibSteer1} and \cite{CopKibSteer2} to consider the kinematic constraints that apply to the formation of $(p,q)$-junctions. As these results are vital for our analysis we briefly review them here.  Considering  the collision of straight strings with charges $(p_1,q_1)$ and $(p_2,q_2)$ meeting at an angle $\alpha$ and traveling with equal and opposite velocities $v$, when the strings collide, they may become linked by a string with charges $(p_3,q_3)=-(p_1+p_2,q_1+q_2)$.  Imposing the requirement that the length of the joining string must increase in time leads to the strong kinematic constraints \cite{CopFirKibSteer}: 

\be\label{f_constr}
f_{\vec\mu}(v,\alpha) \equiv A_1 (1+v^2)^2 +
A_2 (1+v^2) + A_3 < 0 \,,
\ee
where
\bq\label{As}
A_{1} &=&  \bar \mu_{+} ^{2} \cos^{2} \alpha \left[  \bar \mu_{3}^{2} - \bar \mu_{+}^{2} \sin^{2} \alpha 
 - \bar \mu_{-}^{2}   \cos^{2} \alpha  \right], \nonumber\\
A_{2}&=& 2  \bar \mu_{+}^{2}  \bar \mu_{-}^{2}  \cos^{2} \alpha
-    \bar \mu_{3}^{4} - (2  \cos^{2} \alpha -1)   \bar \mu_{+}^{2}  \bar \mu_{3}^{2} ,
  \nonumber\\
A_{3}&=&\bar \mu_{3}^{4} - \bar \mu_{+}^{2}  \bar \mu_{-}^{2}.
\eq
and $\bar \mu_{\pm}= \bar \mu_{1} \pm \bar \mu_{2}$. 

The inequality (\ref{f_constr})  leads to a condition on the values of $v$ for which junction formation is possible, namely:
\be
0 \leq v^2 < v_c^2(\alpha) ,
\ee
where the critical velocity, $v_{c}$, also depends on $\bar \mu_{1}$ and $\bar \mu_{2}$, hence on the charges of the two colliding strings.
In \cite{CopKibSteer2}, it was shown that  $v_c^{\rm max}\leq 1$,  (where $v_c^{\rm max}$ is defined as the maximum value taken by $v_c$ as $\alpha$ varies from $0$ to $\pi/2$),  only if
\be
\bar \mu_3^2 \geq  \bar \mu_{+} |\bar \mu_{-} |=  |\bar \mu_1^2 - \bar \mu_2^2| .
\label{cc}
\ee
Hence, if this  condition is satisfied by the tensions of the joining strings, then there is a velocity $v_c^{\rm max}$ above which the two colliding strings will pass through each other rather than forming a junction.

A couple of examples may help illuminate the effect of the constraints \cite{CopFirKibSteer}. For the collision of equal tension strings (for example (1,1) or (1,-1) strings), then $\bar \mu_{-}=0$ and the constraint Eqn.~\ref{f_constr} becomes 
\be
\sqrt{1-v^2}  \cos\alpha > \frac{\bar \mu_3}{2\bar \mu_1}. 
\label{eq}
\ee
Equation (\ref{eq}) holds as long as the triangle inequalities are satisfied amongst the three tensions, namely that for all $i$
\be
\bar \nu_i \geq 0 
\ee
where, for example,
\be
\bar \nu_1 = \bar \mu_2 + \bar \mu_3 - \bar \mu_1 
\ee
(and cyclic permutations). For $g_s \ll 1$, the largest region of the $(\alpha, v)$ plane is available to the collision of ($p_1,q_1$) with ($p_1,-q_1$) strings which produce a joining string with charges ($-2p_1,0$). Such a string is a $p$-string which is light, indicating that it is much more likely to form a light joining $p$-string than a heavy joining $q$-string. There is a symmetry present which means that if the second string is of the $(-p_1,q_1)$ type, then because of the invariance under $p\rightarrow q$ and $g_{s} \rightarrow 1/g_{s}$ (basically because F and D-strings are symmetric under string theory S-duality where weak coupling is replaced by strong coupling, i.e.  $g_{s} \rightarrow 1/g_{s}$), then the linking string is a D-string.

A second example is the collision of an F, (1,0)-string with a D, (0,1)-string. Such an event is important because it forms the building blocks for general (p,q) string collisions. Of course the third string formed at the junction is a (1,1) string. 

In \cite{CopFirKibSteer}, under the assumption that an $x$-link forms and $2 \cos^{2} \alpha >1$, the condition Eqn.~(\ref{f_constr}) for junction formation is shown to become  $0< v^{2} < v_{c}^{2}$ where 
\be
\label{gmc1}
v_{c}^{2} = \frac{  (1+g_{s}^{2} ) - 4 \cos^{2}  \alpha \sin^{2} \alpha (1+g_{s} )^{2}  
+ \sqrt{  (1+g_{s}^{2} )^{2}   -    4 \cos^{2}  \alpha \sin^{2} \alpha (1-g_{s}^{2} )^{2}   } }{2  \cos^{2}  \alpha  (1+g_{s} )^{2}    (2  \cos^{2} \alpha -1)     } \,.
\ee
One of the interesting features that emerge is $v_c=0$ when $\alpha=\pi/4$, independent of $g_{s}$. For $\alpha > \pi/4$ no $x$-link can be formed in this case, whatever the values of $g_s$ or $v$.

As $g_{s} \rightarrow 0$, in which case $v_{c}=1$ then half the $(\alpha,v)$ plane is allowed implying physically, that a very heavy D-string can always exchange partners with a light F-string, with the third string moving with a velocity approximately equal and in a direction almost parallel to  the incoming heavy D-string. A second limit is when $g_{s} \rightarrow 1$ in which case the F-string is almost as heavy as the D-string and $\bar \mu_{-} \rightarrow 0$, which corresponds to the case described above for equal tension strings. 

It is clear from this brief discussion that the presence of the kinematic constraints has a big effect on the ability of colliding strings to form junctions. We will now incorporate these constraints in the string evolution models and see how they have a dramatic effect on the scaling solutions obtained. 

\section{\label{incorp}Incorporating the Constraints in our String Evolution Models}

The constraints described in the previous section act as to 
``switch off'' certain interaction processes with parameters ($v,\alpha$) outside the kinematically allowed range.  These constraints have
been derived from the Nambu-Goto equations and so they only (strictly)
hold in the thin-string approximation.  However, they have recently
been found to be in excellent agreement with field theory simulations,
based on Abelian-Higgs models~\cite{Salmi,Bevis:2008hg}, although of course there are specific differences arising from the particular nature of the strings and their interactions (see also \cite{Urrestilla:2007yw}). One can think of these constraints as window functions which, for
each interaction process involving certain types of strings, restrict the allowed region of  ($v,\alpha$)-parameter space in which junctions can form.
Thus, in a phenomenological approach like that of section \ref{models},
where only the average effect of string collisions over the whole parameter
range is considered, one should be able to integrate these constraints
over the full range of the parameters ($v,\alpha$)~\cite{thesis,CopKibSteer2}.
This would effectively introduce a suppression in the relevant
coefficients of equations (\ref{rho_idt})-(\ref{v_idt}) that would (in
general) be different for each zipping process, since the constraint
curves depend on the tensions of the strings involved.  The aim of
the present section is precisely to incorporate these constraints
in the network evolution models of section \ref{models}. In particular we will integrate the constraints of section \ref{constrs},
modelled as window functions, over an appropriately normalised probability distribution in ($v,\alpha$)-space.  The result will be a number less than unity, since what we are doing is simply eliminating that part of parameter
space which does not satisfy the constraint.  In other words, we are
effectively switching off a number of kinematically forbidden interactions
that would otherwise be included in the models.  As the string network
is to a good approximation Brownian, the strings are randomly oriented
so we will assume a flat probability distribution for the collision
angles $\alpha$.  On the other hand, the velocities $v$ -- for scaling
networks -- are peaked at a particular value $v_s$, with a Gaussian
probability distribution~\cite{AusCopKib,Polch_priv}.  The variance 
$\sigma_v$ depends on the tensions of the 3 strings involved
(in fact on $\mu_1$ and $\bar\mu\equiv \mu_2-\mu_3$)\footnote{See
equation (64) of~\cite{CopKibSteer2} for $\langle\dot{\bf{x}}^2\rangle=
f(\mu_1,\bar\mu)$.  One can do the same calculation for
$\langle\sqrt{\dot{\bf{x}}^2}\rangle^2$, though in this case the
result cannot be expressed analytically.}, but lies in a narrow
range, for example $\sigma_v^{\,2}\in [0.239,0.275]$ for $0.2<\mu_1<1.4$, $\bar\mu<\mu_1$.  Thus, in the following we will take a universal
variance of $\sigma_v^{\,2}=0.25$ for all interactions.  This can
be easily generalised to $\sigma_v(\vec\mu)$, but the results
will not depend strongly on this choice.

\begin{figure}[h]
  \begin{center}
    \includegraphics[height=1.6in,width=1.8in]{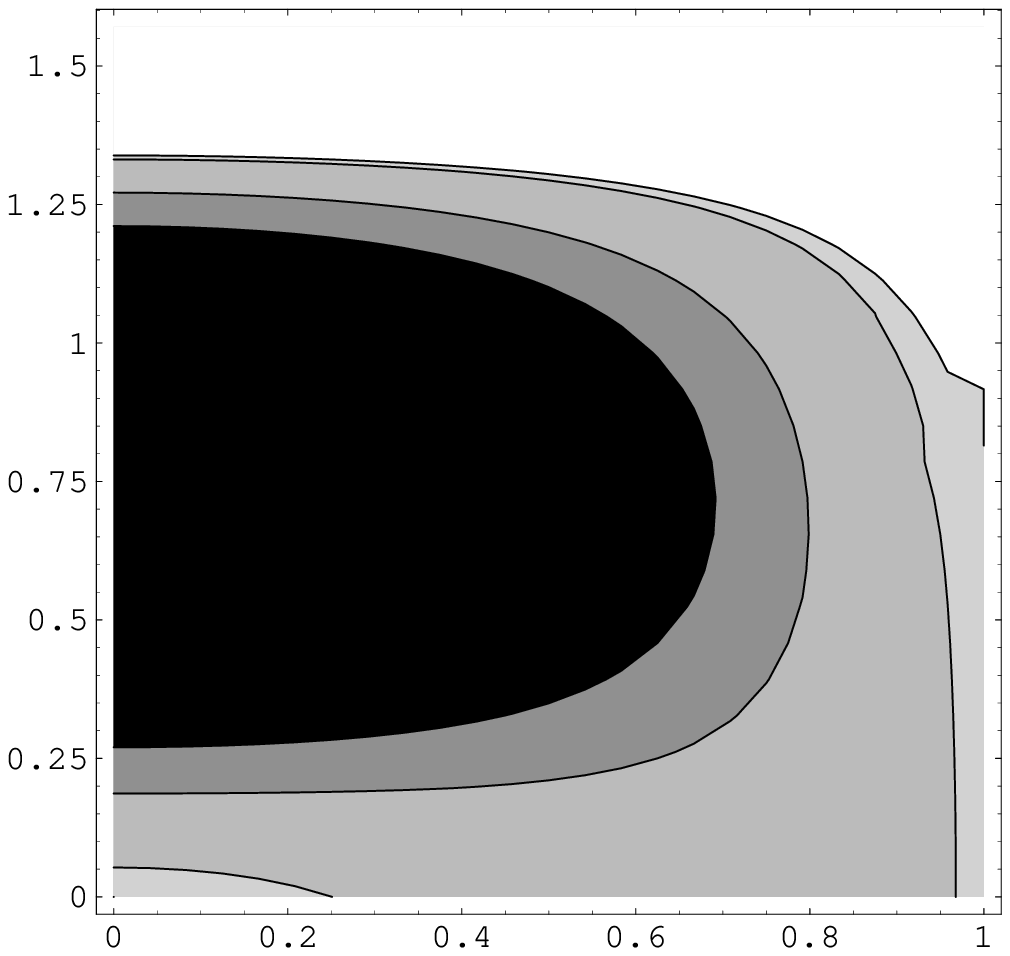}
    \includegraphics[height=1.6in,width=1.8in]{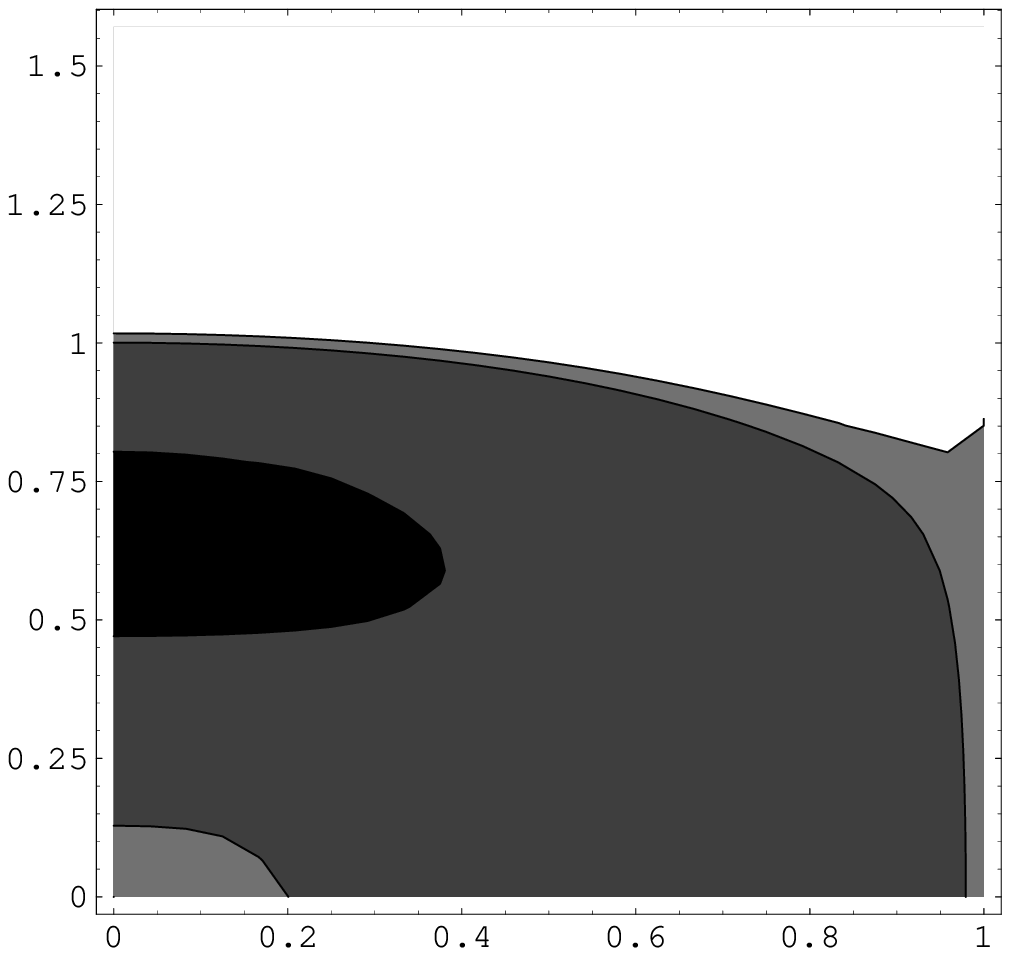}
    \includegraphics[height=1.6in,width=1.8in]{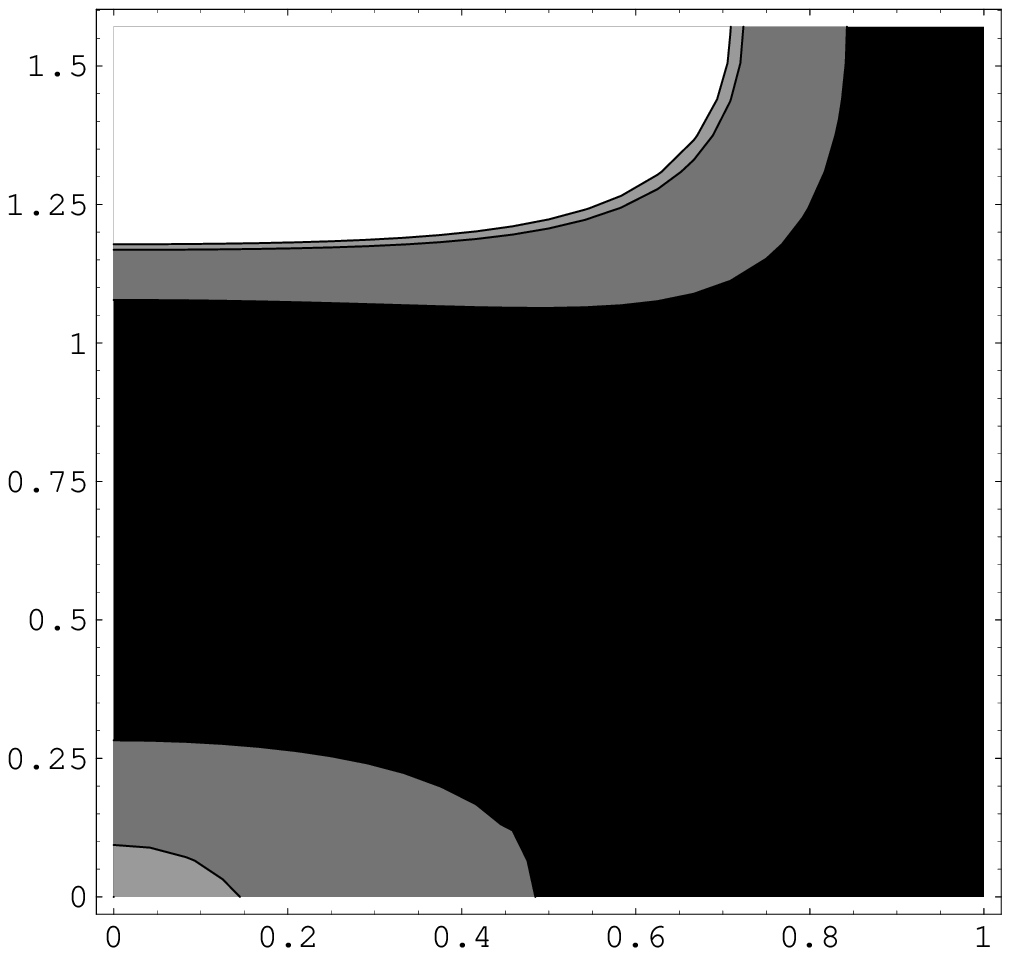}
    \includegraphics[height=1.6in,width=1.8in]{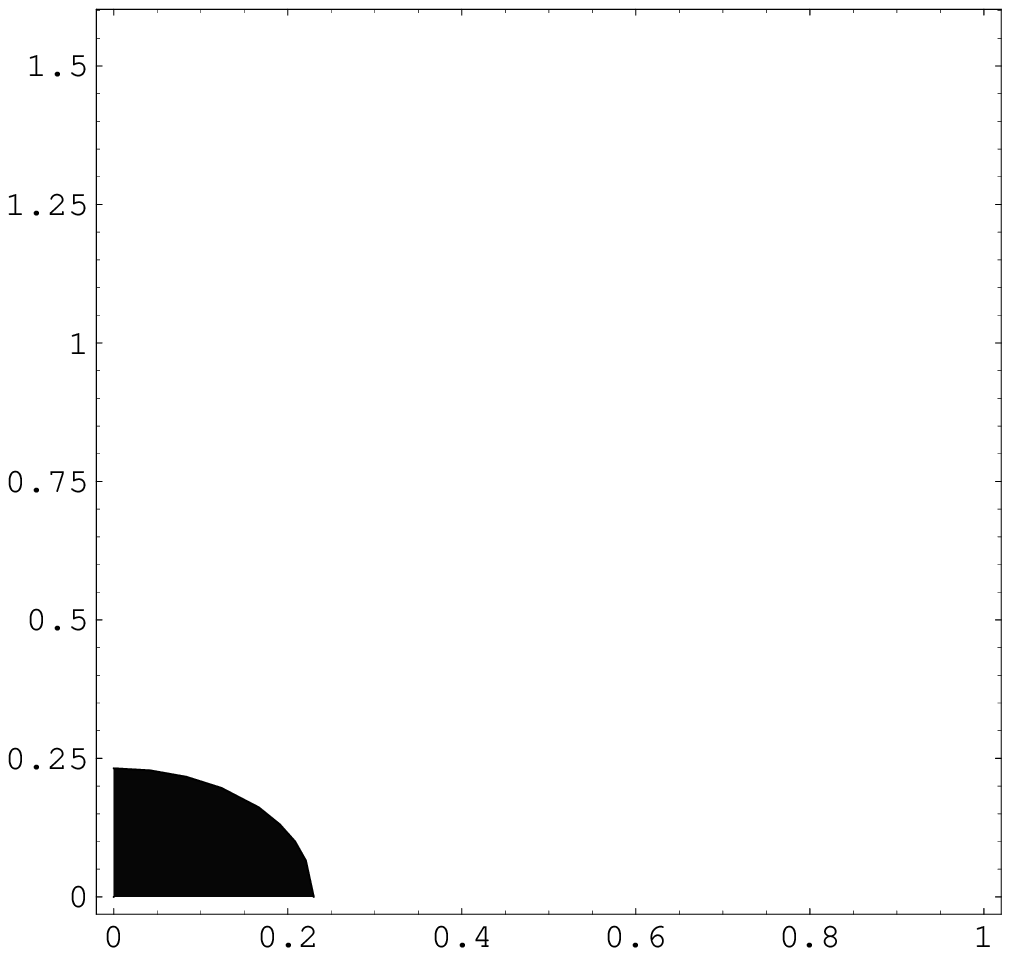}
    \includegraphics[height=1.6in,width=1.8in]{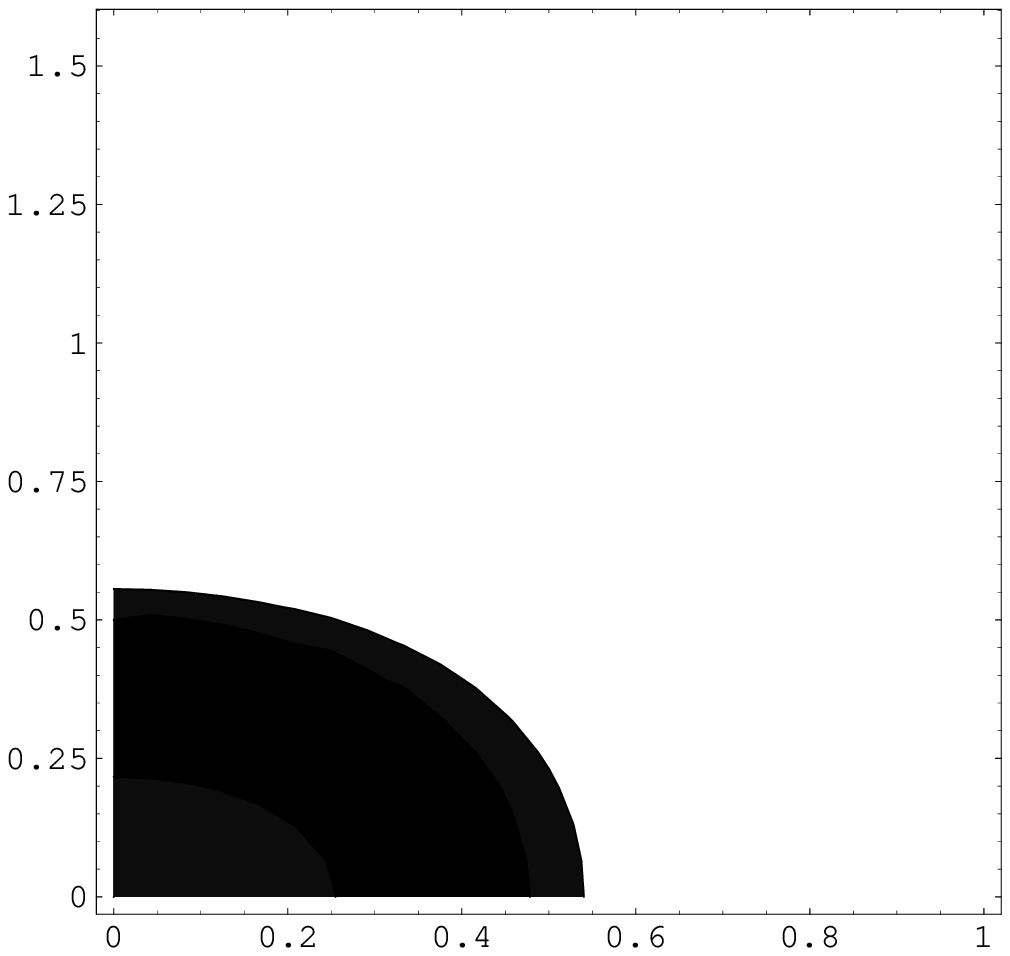}
    \includegraphics[height=1.6in,width=1.8in]{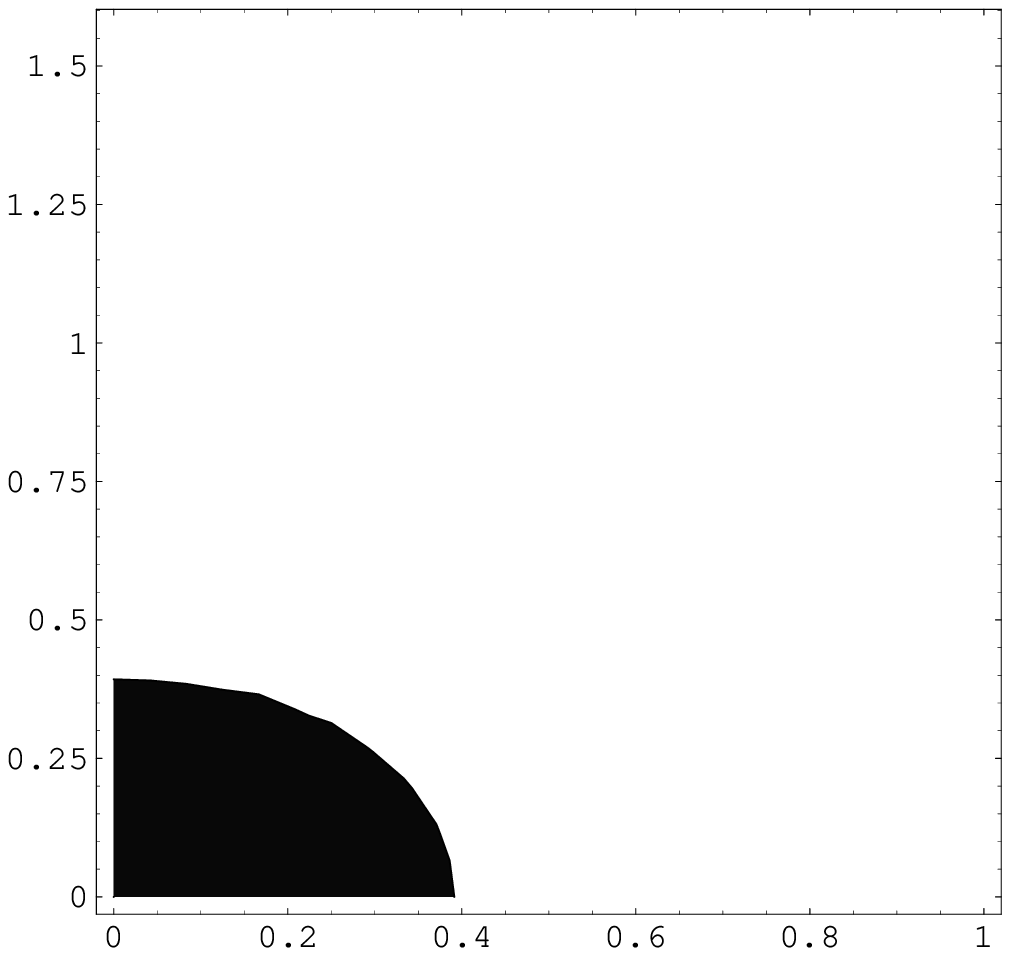} 
    \caption{\label{constr_plots} The allowed parameter range in $(v,\alpha)$-space for different interactions, obtained from Eq.~(\ref{f_constr}).  The interactions shown are (starting from top left): $(0,1)+(1,1)\rightarrow (1,0)$, $(1,0)+(1,1)\rightarrow (0,1)$, $(1,0)+(2,1)\rightarrow (1,1)$, $(0,1)+(1,1)\rightarrow (1,2)$, $(1,0)+(1,1)\rightarrow (2,1)$ and $(1,0)+(2,1)\rightarrow (3,1)$. Upper plots correspond to the subtractive channel (making lighter string states) while lower ones to the additive (making heavier strings).  The area of the allowed region depends strongly on the particular interaction, the subtractive channel (upper plots) being favoured over the additive one (lower plots). 
     }
   \end{center}
  \end{figure}

Therefore, if $f_{\vec\mu}(v,\alpha)$ is the constraint function in
equation (\ref{f_constr}), so that the kinematically allowed region
is given by $f_{\vec\mu}(v,\alpha)<0$, then the relevant suppression
factor for the process where strings of types $i$ and $j$ interact to
form a type $k$ segment will be
\be\label{suppres}
S_{ij}^k = \frac{2}{\pi} \int_0^1 \int_0^{\pi/2}
\Theta(-f_{\vec\mu}(v,\alpha)) \exp[{(v-\bar v_{ij})^2/\sigma_v^{\,2}}]
{\rm d}\alpha {\rm d}v < 1 \,,
\ee
where $\Theta(x)$ is the Heaviside (Unit Step) function and $\bar v_{ij}$ is the RMS relative velocity between strings of types $i$ and $j$, that is 
\be\label{vbar}
\bar v_{ij}\equiv
\sqrt{\left\langle[{\bf v}_i(t)-{\bf v}_j(t)]^2 \right\rangle}=\sqrt{v_i(t)^2+v_j(t)^2}\,.
\ee
Thus, the suppression factor (\ref{suppres}) depends on the particular interaction process through the tension dependence of $f_{\vec\mu}(v,\alpha)$
and the relative velocity $\bar v_{ij}$\footnote{Also through the variance if the tension dependence of $\sigma_v$ is taken into account.}.  As shown in Fig.~\ref{constr_plots} these suppression factors are numerically very different for different interactions, so it is important to take them into account in macroscopic evolution models.  

The constraints can be then implemented in the (multi-)network evolution
models of section \ref{models} by the replacement:
\be\label{new_d}
\tilde d_{ij}^k\rightarrow \tilde D_{ij}^k\equiv S_{ij}^k \, \tilde d_{ij}^k \,.
\ee
Clearly, with this replacement, the coefficients of the 3-string interaction
terms in equations (\ref{rho_idt})-(\ref{v_idt}) depend non-trivially on
the unknown functions $v_i(t)$ through the ``parameter'' $\bar v_{ij}$
(see equations (\ref{suppres}) and (\ref{vbar})), so the system is
numerically heavy to integrate.  However, for scaling networks, one
can solve the system iteratively, treating $\bar v_{ij}$'s as constant parameters in each iteration.  In the first step, one solves the system
with initial guess-values assigned to $\bar v_{ij}$'s, and then solves
the system again, replacing the initial guesses for $\bar v_{ij}$'s with
the corresponding values obtained by inserting the returned scaling values
of $v_i(t)$'s into equation (\ref{vbar}).  Repeating this procedure (in
practice two or three iterations are needed), one can match the input
$\bar v_{ij}$'s to those obtained at scaling, to the required accuracy.
In the next section we follow the above procedure to solve the system in
the case of cosmic superstrings for different sets of values of $\tilde c_i$'s and $\tilde d_{ij}$'s.

\section{\label{results}Network Results}

The general behaviour of multi-tension string networks -- without taking into account the above kinematic constraints -- has been investigated by a number of authors in different contexts \cite{McGraw,SpergPen,TWW,CopSaf,NAVOS,HindSaf,Saffin:2005cs,Rajantie:2007hp,Urrestilla:2007yw} and in particular, networks modelled by equations (\ref{rho_idt})-(\ref{v_idt}) have been studied in~\cite{NAVOS}.  Here we will concentrate on the effects of the kinematic constraints of section~\ref{constrs}, which affect the network dynamics  through the fact that they rescale the $\tilde d_{ij}^k$ terms (\ref{new_d}).       

Let us first consider a simple model in which the interaction terms in the velocity evolution equations (\ref{rho_idt}) are switched off, $\tilde b_{ij}^k=0$.  Physically, this corresponds to the assumption that, during the binding process between strings of type $i$ and $j$, forming a (lighter) string segment of type $k$, the energy difference arising from the disparity in string tensions is efficiently radiated away.  This may be plausible for cosmic superstring networks, where binding is accompanied by the production of microscopic F-strings that can damp energy away.  It was pointed out in reference~\cite{TWW} that this mechanism alone can be sufficient to drive the network to a scaling regime, that is, with the assumption that string binding dissipates all the energy imbalance associated with junction formation processes, such junction-forming multi-tension networks can reach a scaling regime even when self-interactions among segments of the same type (which for single string networks provide the dominant energy-loss mechanism through loop production \cite{book}) are ignored.  

The system (\ref{rho_idt})-(\ref{v_idt}) with the interaction rules (\ref{Paddsubtr}) - (\ref{d_super}), and $\tilde b_{ij}^k=0$ generalises the model of~\cite{TWW} in that it assigns a different correlation length and a different RMS velocity for each of the network components, as described in section \ref{models}.  Even in this more general class of models the main result of \cite{TWW} remains true and scaling is generically reached, with heavier string components being systematically less populated.  This is because the factors $P_{ij}^\pm$ in the coefficients $\tilde d_{ij}^k$ of  (\ref{rho_idt}) favour the subtractive over the additive interaction channel, so lighter strings acquire larger number densities compared to heavier ones.  However, the relative abundance among strings of different types depends quantitatively on the numerical values of $\tilde c_i$ -- parametrising loop production -- and $\tilde d_{ij}=\tilde d_{ij}^k/P_{ij}^\pm$ -- corresponding to binding interactions.  Therefore, the kinematic constraints on string junctions, described in the previous section, can significantly affect the relative abundances of the most populated, lighter network components through the suppression factors (\ref{suppres}),(\ref{new_d}), which should therefore be included in analytic models of this type.  

\begin{figure}[h]
  \begin{center}
    \includegraphics[height=2.1in,width=2.4in]{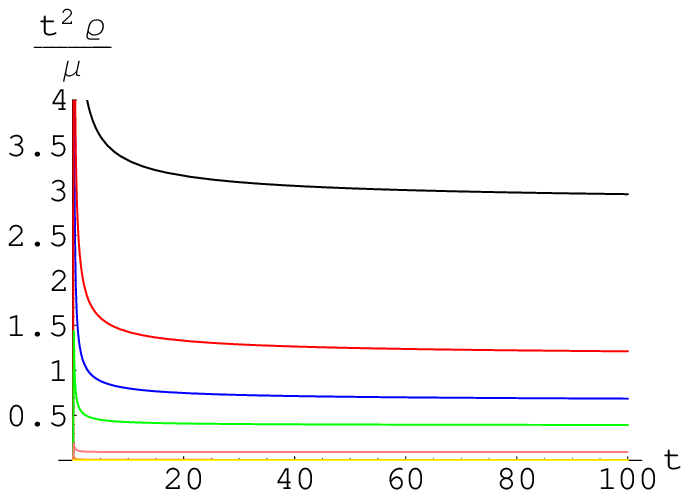}
    \includegraphics[height=2.1in,width=2.4in]{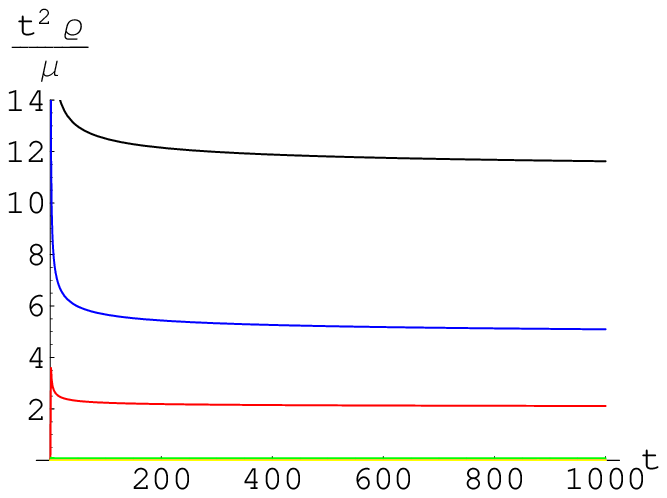}
    \includegraphics[height=2.1in,width=2.4in]{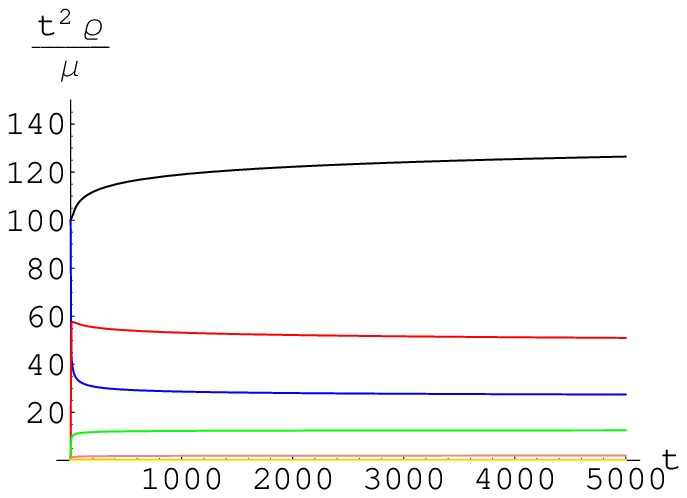}
    \includegraphics[height=2.1in,width=2.4in]{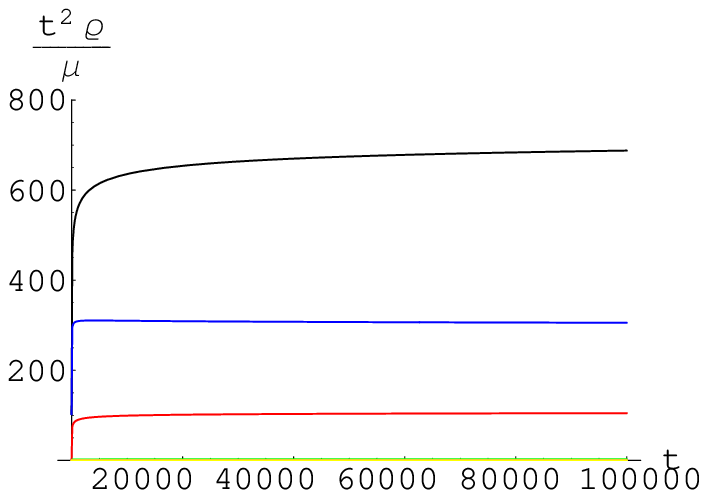}
    \caption{\label{rhofigs} Evolution of normalised string density
             $N=\rho t^2/\mu$ in the radiation era for the lightest components 
             of a network described by equations  
             (\ref{rho_idt}),(\ref{v_idt}),(\ref{Paddsubtr}),(\ref{d_super})
             with $\tilde b_{ij}^k=0$ and for $g_s=0.3$.  F-strings are shown 
             in black, D-strings in blue, $(1,\pm 1)$ strings in red, 
             $(2,\pm 1)$ in green, $(1,\pm 2)$ in orange, $(3,\pm 1)$ in 
             pink and $(1,\pm 3)$ in yellow.  Plots on the left do not take 
             into account the effects of junction constraints, while in the 
             plots on the right, these effects are included through the 
             coefficient rescalings (\ref{new_d}).  Upper plots are for 
             $\tilde c_i=0.23$, $\tilde d_{ij}=1$, while lower plots for
             $\tilde c_i=0.023$, $\tilde d_{ij}=0.1$.}
   \end{center}
  \end{figure}

Figure \ref{rhofigs} shows the scaling densities of such a network with string coupling $g_s\!=\!0.3$, for $\tilde c_i=0.23$\footnote{This value corresponds to that obtained by fitting Nambu-Goto simulations of single string networks \cite{VOS}.}, $\tilde d_{ij}=1$\footnote{The $\tilde d_{ij}$'s are chosen to be significantly larger than the $\tilde c_i$'s to demonstrate that, for our $\tilde b_{ij}^k=0$ model which is analogous to that of reference \cite{TWW}, scaling can be found even though the junction formation terms are dominant.  This is to be contrasted to the energy-conserving model $\tilde b_{ij}^k=\tilde d_{ij}^k$, where scaling cannot be achieved for such large values of $\tilde d_{ij}$.} (top panel) and for $\tilde c_i=0.023$, $\tilde d_{ij}=0.1$ (bottom panel), in the radiation era.  Although the results here are for the radiation era, similar conclusions hold in the matter era but with different specific values for the scaling densities and parameters $\tilde c_i$ and $\tilde d_{ij}$.  The formally infinite set of equations (\ref{rho_idt})-(\ref{v_idt}) has been truncated at 4-string composites, leaving only seven species of distinct tension, namely (1,0) or F-strings, (0,1) or D-strings, $(\pm 1,1)$, $(2,\pm 1)$, $(\pm 1,2)$, $(3,\pm 1)$ and $(\pm 1,3)$, shown in black, blue, red, green, orange, pink and yellow respectively.  The plots on the left have been obtained without taking into account the junction constraints of sections \ref{constrs} and \ref{incorp}, while those on the right demonstrate the effect including these constraints have on the string scaling densities.  Note on the left panel that, for these models $(\tilde b_{ij}^k=0)$, scaling is reached even though the cross-interaction coefficients $\tilde d_{ij}$ are significantly larger than the self-interaction ones.  Also note, that if we turn off the parameter determining the efficiency of cross-interactions, (i.e. we set  $\tilde d_{ij}=0$), then a network forms which is a tangle of different string species with tensions given by (\ref{tensions}) that do not interact with one another. Such a network would reach a scaling regime in which all network components independently scale due to their self-interactions parametrized by $\tilde c_i$.  For $\tilde c_i=0.23$ (all $i$) all network components would asymptotically reach a normalised string density of $N\equiv t^2 \rho_i / \mu_i\simeq 14$  in the radiation era, while for $\tilde c_i=0.023$ the corresponding value would be $N \simeq 980$\footnote{This enhancement of string density resulting from a suppression of string self-interactions has generated significant interest in networks of cosmic superstrings, because their intercommutation probabilities can be suppressed compared to field theory strings~\cite{JST,JackJoPolch}.  This in effect reduces $\tilde c_i$ so cosmic superstring networks may have much larger number densities than their field theoretic counterparts.  However, numerical simulations~\cite{intProb,Vanchurin} suggest that this effect is not as strong as originally anticipated.}.  

The overall reduction of scaling densities in the left panel with respect to the corresponding values for $\tilde d_{ij}=0$ just mentioned, reflects the impact of choosing values of $\tilde d_{ij} > \tilde c_i$. In that case the dynamics is dominated by energy losses associated with junction forming binding processes. If we also assume that the energy can be radiated away, then the energy loss due to binding can be greater than the corresponding losses to loop-production.  As noted in reference~\cite{NAVOS} -- and as will be demonstrated below -- this is not observed in the case of energy conserving models, $\tilde b_{ij}^k=\tilde d_{ij}^k$, where the energy liberated by forming junctions is redistributed in the network through momentum exchange: such networks simply fail to reach scaling for large $\tilde d_{ij}$.  The important effect of cross-interactions $\tilde d_{ij}\ne 0$, subject to quantitative modifications due to the inclusion of the junction constraints (right panel), is to suppress the abundance of heavy species relative to light ones through the `selection rule' (\ref{Paddsubtr}), which strongly favours the subtractive interaction channel forming lighter strings.        

A comparison between the left and right panels of Fig. \ref{rhofigs}, shows that taking into account the kinematic constraints of equations (\ref{f_constr})-(\ref{As}) is crucial for making quantitative predictions.  By removing kinematically forbidden processes that would otherwise have been included, it introduces both an overall suppression factor with respect to self-interactions (which are not subject to these constraints) and a relative weighting among different cross-interactions.  The overall suppression makes self-interactions more important, thus driving the average string densities closer to their self-interacting values ($N\simeq 14$ for $\tilde c=0.23$, $N\simeq 980$ for $\tilde c=0.023$), while the relative weighting changes the abundances among different string species.  The general trend that heavy, composite strings are systematically suppressed is observed,  the dominant species being F-, D- and $(\pm 1,1)$-strings.      

\begin{figure}[h]
  \begin{center}
    \includegraphics[height=2.1in,width=2.4in]{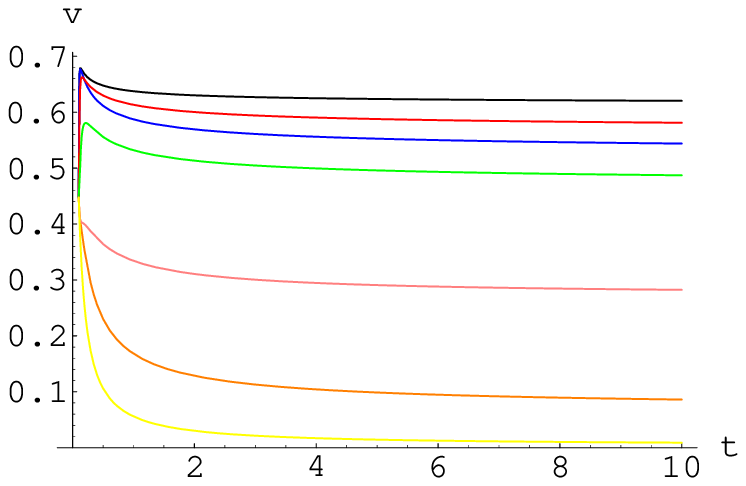}
    \includegraphics[height=2.1in,width=2.4in]{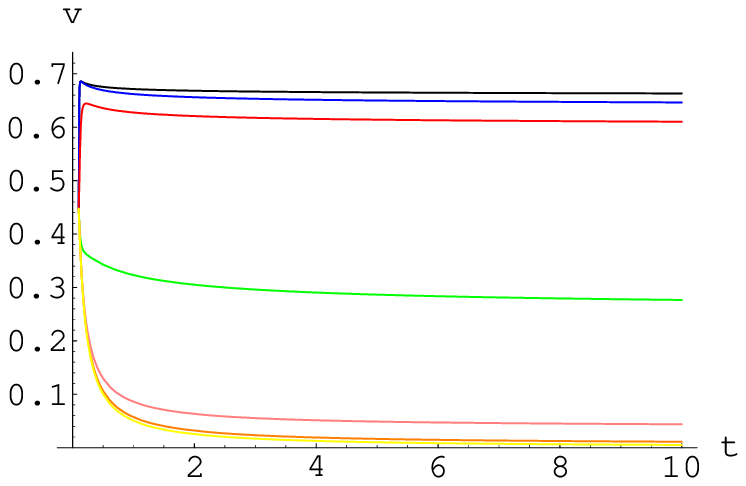}
    \includegraphics[height=2.1in,width=2.4in]{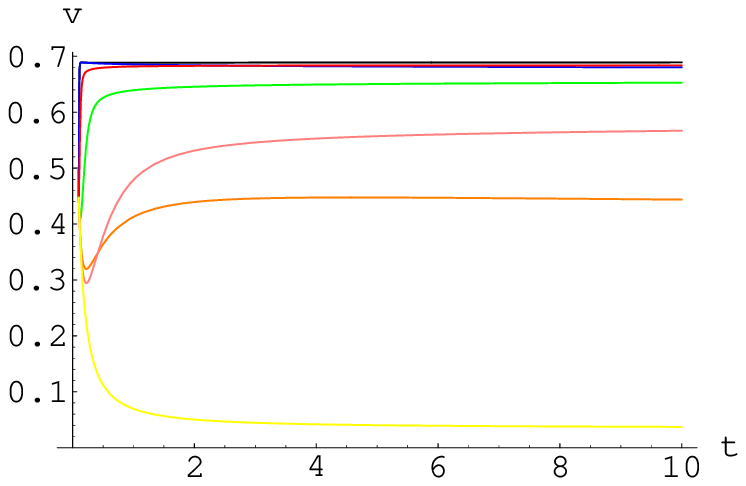}
    \includegraphics[height=2.1in,width=2.4in]{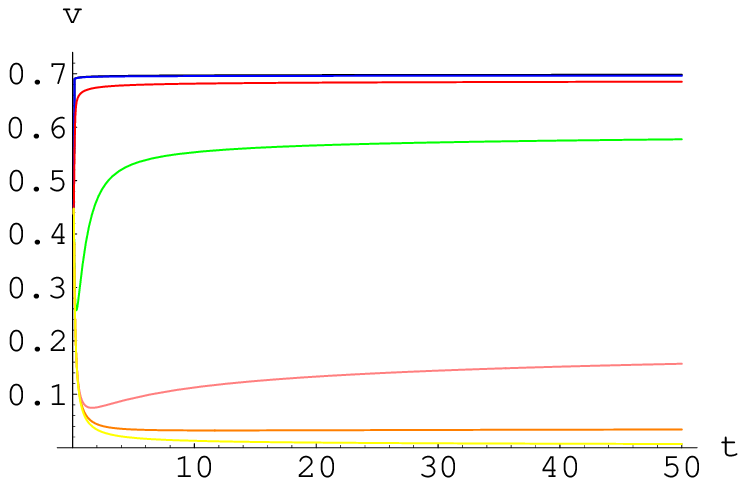}
    \caption{\label{vfigs} Velocity evolution for the string networks of 
             Fig.~\ref{rhofigs} ($\tilde b_{ij}^k=0$, $g_s=0.3$). 
             F-strings are shown in black, D-strings in blue, $(1,\pm 1)$
             strings in red, $(2,\pm 1)$ in green, $(1,\pm 2)$ in orange, 
             $(3,\pm 1)$ in pink and $(1,\pm 3)$ in yellow.  Plots on the 
             left do not take into account the effects of junction 
             constraints, while those on right do; upper plots are for
             $\tilde c_i=0.23$, $\tilde d_{ij}=1$, and lower plots for 
             $\tilde c_i=0.023$, $\tilde d_{ij}=0.1$.}
   \end{center}
  \end{figure}

The corresponding scaling solutions for the network RMS velocities are shown in Fig. \ref{vfigs}.  Heavier, less populated strings are systematically slower, with the heaviest ones having negligible RMS velocities as their correlation length falls outside the horizon.  This effect is exaggerated in these simplified models $(\tilde b_{ij}^{k}=0)$, which ignore momentum transfers during binding processes; equations (\ref{v_idt}) for $\tilde b_{ij}^k=0$ only take into account correlation-length-scale string curvature and Hubble damping, so, once a string species correlation length exits the horizon, Hubble damping dominates.  At smaller scales, the velocity is controlled by short-scale structure, which is not captured in detail by these models.  Allowing momentum transfers among different networks, introduces an extra source that can significantly affect the velocities of heavy strings, as will be demonstrated later. The basic feature that the heavier strings are systematically slower is consistent with the  analytic analysis of  the RMS velocities of junctions performed in \cite{CopKibSteer2,CopFirKibSteer}.       

\begin{figure}[h]
  \begin{center}
    \includegraphics[height=2.1in,width=2.4in]{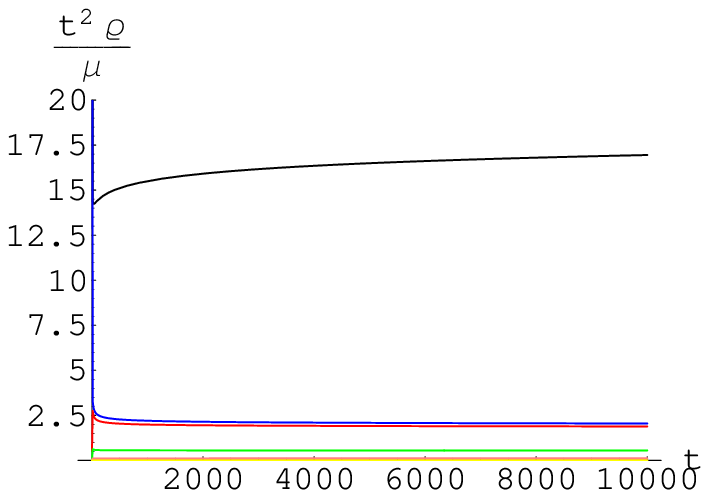}
    \includegraphics[height=2.1in,width=2.4in]{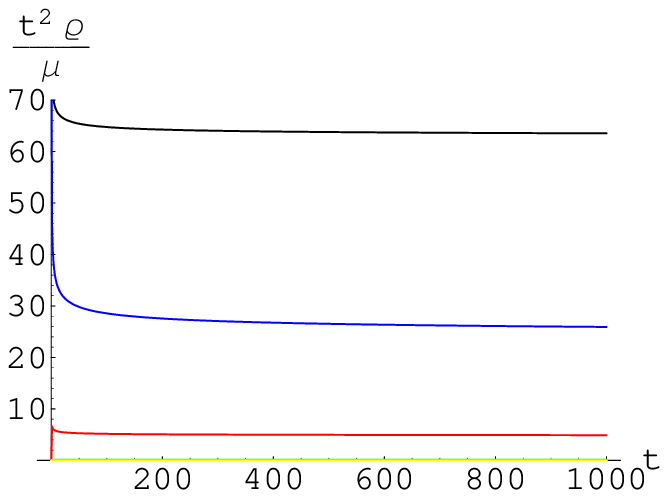}
    \caption{\label{figs_diffPs} Evolution of normalised string density
             $N=\rho t^2/\mu$ in the radiation era, for a network with 
             $\tilde b_{ij}^k=0$, but with a more realistic choice of the 
             parameters $\tilde c_i$ and $\tilde d_{ij}^k$ ($g_s=0.3$).
             Again, F-strings are shown in black, D-strings in blue, 
             $(1,\pm 1)$ strings in red, $(2,\pm 1)$ in green, $(1,\pm 2)$ 
             in orange, $(3,\pm 1)$ in pink and $(1,\pm 3)$ in yellow.  
             Here, $\tilde c_i=0.23 \times P_i^{1/3}$ where, in the above 
             order, $P_i=(0.05, 0.1, 0.6, 0.7, 0.9, 0.8, 1)$.  The power 
             $1/3$ comes from the numerical simulations of 
             Ref.~\cite{intProb}. The coefficients $\tilde d_{ij}$ are: $\tilde d_{12}=0.4, \tilde d_{13}=0.6, \tilde d_{14}=0.65,\tilde d_{15}=0.8, \tilde d_{16}=0.75, \tilde d_{17}=0.95, \tilde d_{23}=0.5, \tilde d_{24}=0.75, \tilde d_{25}=0.85, \tilde d_{26}=0.8, \tilde d_{27}=0.95, \tilde d_{34}=0.75, \tilde d_{35}=0.87, \tilde d_{36}=0.8, \tilde d_{37}=0.9, \tilde d_{45}=0.9, \tilde d_{46}=0.9, \tilde d_{47}=0.95, \tilde d_{56}=0.9, \tilde d_{57}=0.95$ and $\tilde d_{78}=0.95$ (all unrelated are null). 
             Plots on the right (left) do (do not) take into account the 
             kinematic constraints on junctions.}
   \end{center}
  \end{figure}

In Figs. \ref{rhofigs} and \ref{vfigs} we have assumed that all self-interaction and cross-interaction processes have the same efficiency for all network components, that is, we have chosen a single value for all $\tilde c_i$'s (the efficiency with which strings of type $i$ self-interact producing loops) and a single value for all $\tilde d_{ij}$'s (the efficiency with which strings of type $i$ bind with strings of type $j$).  This was only to demonstrate the general effects of cross-interactions and of junction constraints on the macroscopic properties of the network.  In principle, the self-interaction and cross-interaction efficiencies depend on the types of strings involved and are related to the quantum intercommuting probabilities of each process, which can be estimated using string theory methods~\cite{JackJoPolch,Jackson}.  These are strongly model-dependent, but the general picture resulting from such analysis is that intercommuting probabilities for F-strings are likely to be in the range $P_{\rm FF}\in[0.01,0.1]$ while for D-strings $P_{\rm DD}\in[0.1,1]$.  For interactions between F- and $(p,q)$-strings, the corresponding probabilities are enhanced with respect to $P_{\rm FF}$ by at least an inverse factor $g_s$, while for interactions between heavy composites one expects probabilities of order unity due to large multiplicities of the relevant Chan-Paton factors as well as due to a suppression of fluctuations for heavier strings.  These different intercommuting probabilities must be reflected in the coefficients $\tilde c_i$, $\tilde d_{ij}$ (although in a non-linear way~\cite{intProb}]) and can crucially affect the relative string abundances.  In particular, the generic situation is $P_{\rm FF}<P_{\rm DD}\Rightarrow \tilde c_{\rm F}<\tilde c_{\rm D}$ so these networks are generally dominated by the lightest $(1,0)$ F-string components.  It is worth therefore solving the system for somewhat more realistic values of the relevant parameters (Fig. \ref{figs_diffPs}).  The assumed values for the parameters $\tilde c_i$ and $\tilde d_{ij}$ are shown in the legend. 

As already mentioned, the simple model described above is somewhat oversimplified in that, by switching off the interaction terms in the velocity equations (\ref{v_idt}), it neglects the effect of momentum transfer among different network components on the RMS velocity of each string type, which in turn affects string densities through the velocity dependencies of equations (\ref{rho_idt}).  Furthermore, the assumption that the energy differences associated with string binding can be efficiently radiated away, critically affects the cosmological evolution of the network.  Indeed, if this assumption did not strictly hold, the simple system considered would artificially damp energy away leading to a possibly `spurious' scaling regime.  A more conservative approach would be to enforce energy conservation during string interactions by redistributing the relevant energy differences within the network through momentum transfer among interacting string segments.  As mentioned earlier this is described by including the interaction terms in the velocity evolution equations (\ref{v_idt}) with $\tilde b_{ij}^k=\tilde d_{ij}^k$.   

In~\cite{NAVOS} it was found that, for this system, scaling is not reached for generic choices of parameters.  The energy liberated by binding processes remains within the network in the form of kinetic energy of newly formed segments, so in this case, and unlike~\cite{TWW}, junction-forming processes alone can never lead to scaling behaviour.  As in the case of single string networks, self-interactions and the related loop production that removes energy from the long-string network, are crucial for scaling to have a chance to develop.  Nevertheless, scaling networks can be readily obtained in this case also, as long as the coefficients $\tilde d_{ij}$ (binding processes) are somewhat smaller than the corresponding loop-production coefficients $\tilde c_i$.  It is clear therefore that, in this case, including the effect of the constraints of section \ref{constrs} in the full system (\ref{rho_idt})-(\ref{v_idt}),(\ref{Paddsubtr}),(\ref{d_super}), with $\tilde b_{ij}^k=\tilde d_{ij}^k$, can make the difference between network frustration and scaling.  In particular, the constraints induce an extra suppression in the coefficients $\tilde d_{ij}^k$ through the redefinition (\ref{new_d}), thereby providing a natural way to implement the requirement of $\tilde d_{ij}^k<\tilde c_i$, as we saw in section \ref{incorp}, a result which favours scaling.       

\begin{figure}[h]
  \begin{center}
    \includegraphics[height=2.1in,width=2.4in]{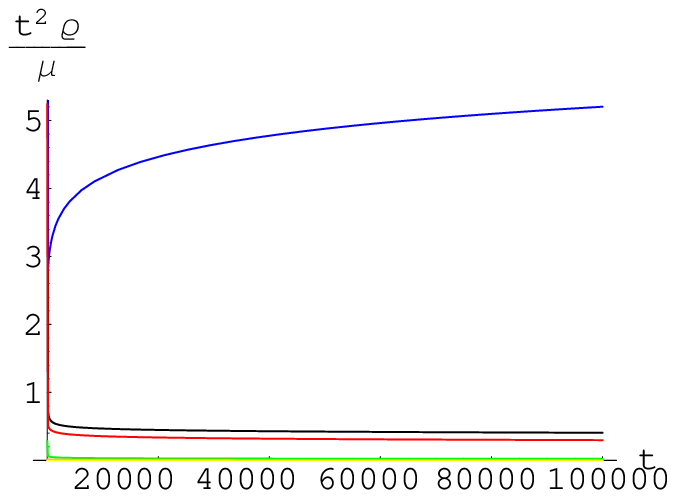}
    \includegraphics[height=2.1in,width=2.4in]{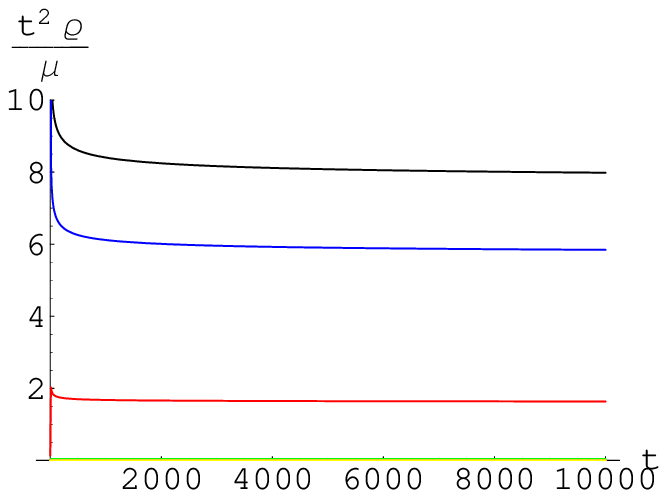}
    \includegraphics[height=2.1in,width=2.4in]{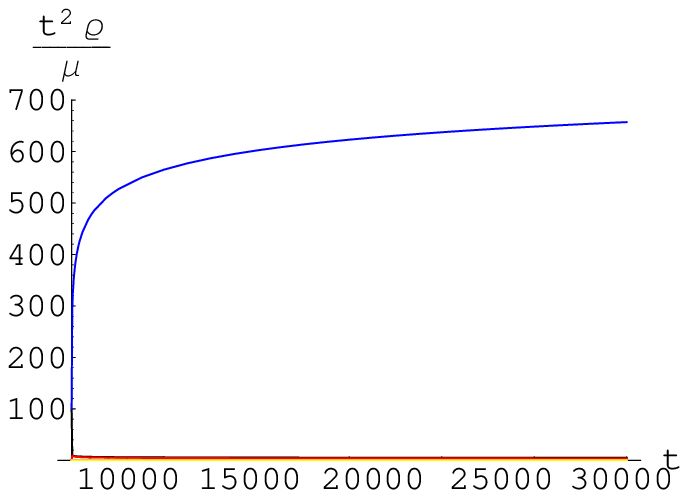}
    \includegraphics[height=2.1in,width=2.4in]{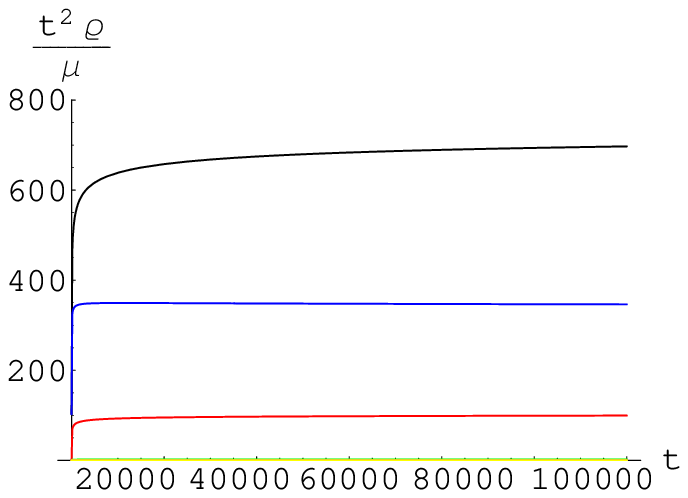}
    \caption{\label{rhofigs_v} Normalised string densities, as in 
    Fig.~\ref{rhofigs}, but now enforcing energy conservation at junction
    formation $\tilde b_{ij}^k=\tilde d_{ij}^k$.  On the left, where the
    effect of the constraints is not taken into account, the networks fail 
    to scale as the junction formation terms dominate.  Including the
    constraints (right) effectively suppresses these terms and scaling
    solutions are observed.}
   \end{center}
  \end{figure}

\begin{figure}[h]
  \begin{center}
    \includegraphics[height=2.1in,width=2.4in]{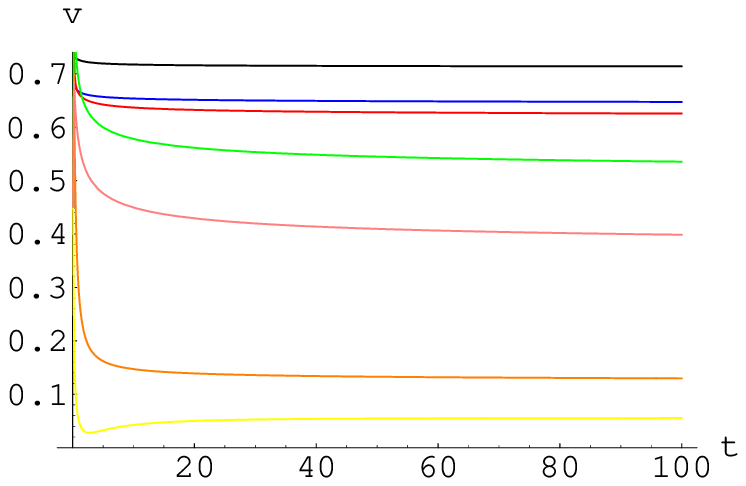}
    \includegraphics[height=2.1in,width=2.4in]{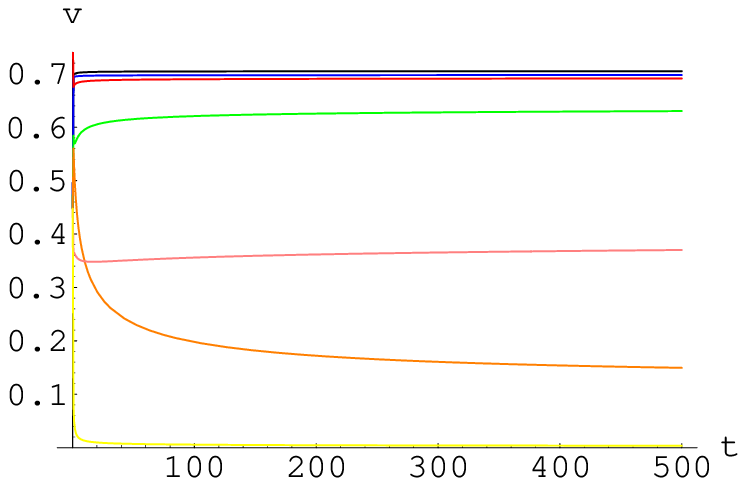}
    \caption{\label{vfigs_v} Network velocities corresponding to the 
    scaling plots (right panel) of Fig.~\ref{rhofigs_v}.}
   \end{center}
  \end{figure}

Fig.~\ref{rhofigs_v} shows the normalised string densities corresponding to Fig.~\ref{rhofigs} but this time for $\tilde b_{ij}^k=\tilde d_{ij}^k$ instead of $\tilde b_{ij}^k=0$.   On the left panel (upper: $\tilde c_i=0.23$, $\tilde d_{ij}=1$; lower: $\tilde c_i=0.023$, $\tilde d_{ij}=0.1$) string densities fail to reach scaling, as the energy liberated by binding processes, which exceeds the energy damped by loop-production, is redistributed within the network as kinetic energy. 
The suppression in the coefficients $\tilde d_{ij}^k=\tilde b_{ij}^k$ due to the junction constraints readily solves this problem, leading to a scaling solution as can be seen in the right panels.  Comparing the right panels of Figs. \ref{rhofigs_v} and \ref{rhofigs} it is clear that, after taking into account the kinematic constraints on string junctions, the effect of momentum transfer can be quantitatively significant, though the results are similar in both cases. The scaling network velocities corresponding to the right panel of Fig.~\ref{vfigs} are shown in Fig.~\ref{vfigs_v}.  Again, heavier, less populated species are slower, but as discussed above the effect is exaggerated in Fig.~\ref{vfigs}.        

\begin{figure}[h]
  \begin{center}
  \includegraphics[height=2.1in,width=2.4in]{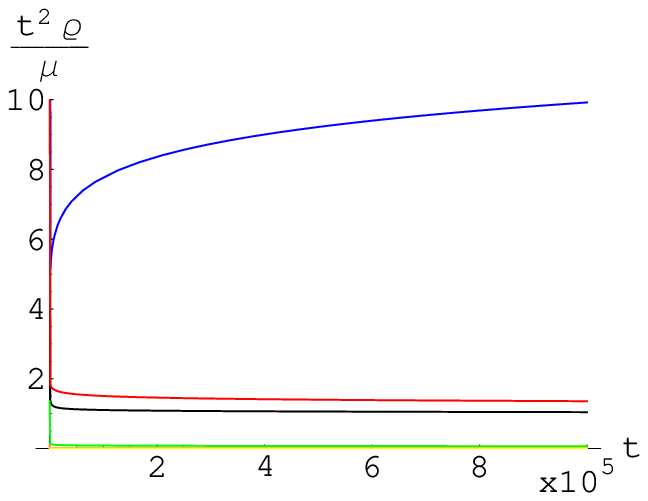}
  \includegraphics[height=2.1in,width=2.4in]{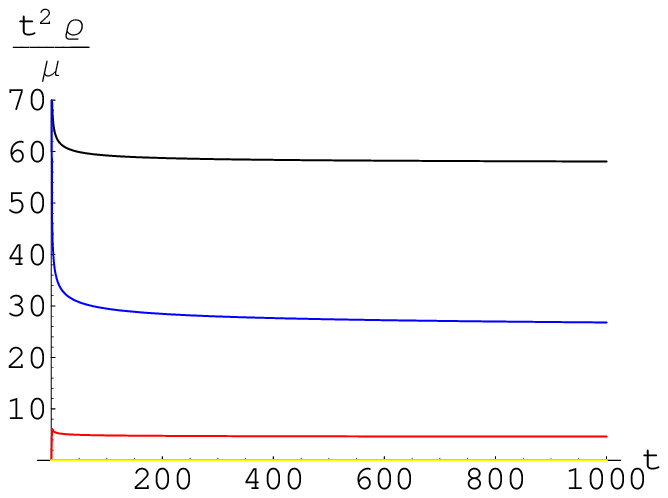}
  \includegraphics[height=2.1in,width=2.4in]{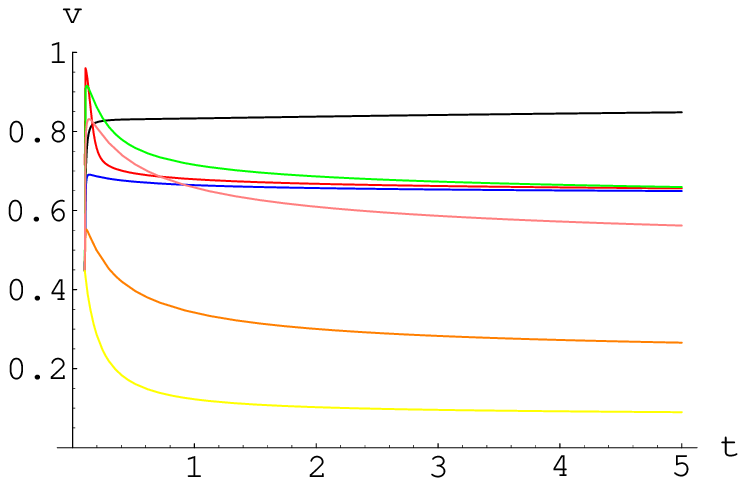}
  \includegraphics[height=2.1in,width=2.4in]{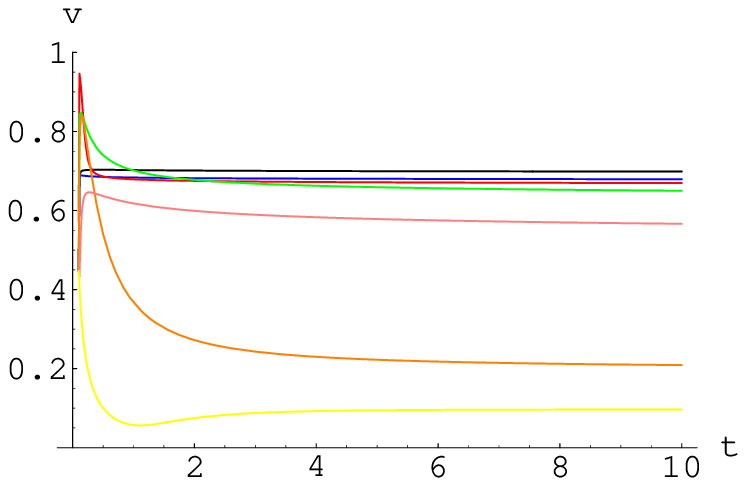}
  \caption{\label{figs_v_diffPs} Normalised string density (upper plots) and 
  network velocities (lower plots) for the parameters of Figs.~\ref{rhofigs} 
  \& \ref{vfigs} but for $\tilde b_{ij}^k=\tilde d_{ij}^k$, enforcing energy 
  conservation at junction formation.  Again, left (right) plots are without 
  (with) the constraints.  As was also the case in Fig.~\ref{rhofigs_v}, 
  these networks do not scale when the constraints are ignored (left).}
  \end{center}
 \end{figure}

Finally, let us consider somewhat more realistic values for the coefficients 
$\tilde c_i$, $\tilde d_{ij}$, as in Fig.~\ref{figs_diffPs}, but now for $\tilde b_{ij}^k=\tilde d_{ij}^k$ instead of $\tilde b_{ij}^k=0$.  This is shown in Fig. \ref{figs_v_diffPs}.  The resulting scaling densities for the chosen intercommuting probabilities are very similar to those of Fig. \ref{figs_diffPs}.  The network is dominated by F-, D-, and $(\pm 1,1)$-strings, in decreasing abundances, with a negligible contribution of heavier species.  Note that the relative abundance of these three lightest species depends on the corresponding intercommuting probabilities, which are model-dependent.  Overall, there is a significant enhancement in string density compared to field theory strings, whose normalised string density would scale at a value of $N\simeq 14$ in the radiation era.   

Now, all the plots in this section are for a string coupling $g_s=0.3$.  For smaller values of $g_s$ the difference in tension between F- and D- strings becomes more pronounced and there is a larger energy gain associated with zipping.  The coefficients $P_{ij}^\pm$ favour even more strongly the `unzipping' of heavier species into lighter ones, and the network becomes more strongly dominated by the lightest F-strings.  For $g_s=1$, on the other hand, F- and D- strings have the same tension and $(p,q)$-, $(q,p)$-strings are degenerate.  There are then only four (instead of seven, in this truncation) non-degenerate tension species whose abundances again fall with increasing tension.  Also, the plots produced here are for evolution in the radiation era.  Network evolution in the matter era reaches scaling at lower string densities (e.g. $N\simeq 3$ for $\tilde c=0.23$, $N\simeq 35$ for $\tilde c=0.023$, both in the self-interacting case $\tilde d=0$).

\section{Discussion}\label{discuss}

Understanding the evolution of a network of cosmic superstrings is a particularly important element in determining the cosmological properties of these intriguing objects. Through a combination of analytical and numerical approaches, evidence has been provided that such complicated networks which involve strings of many tensions which can form three-string junctions can reach scaling solutions, where the energy density in each type of string scales with the background energy density.  In that regime, the lighter strings have a higher density than the heavier strings. In this paper we have extended the semi analytic model of \cite{NAVOS} to include the kinetic constraints obtained in \cite{CopFirKibSteer} for ($p,q$)-strings. These constraints have the effect of limiting the range of relative velocities and angles of approach which can lead to junction formation for  interacting strings. Earlier analysis of the dynamics of string networks have failed to take these influential constraints into account, and in this paper we have established how important they are. In many ways, the nicest aspect of the analysis is that the net effect of the constraints can be accommodated by a rescaling of a single term in the one-scale evolution equations as seen in equation (\ref{new_d}) where the 
effective interaction strength between colliding strings is reduced. This has the knock on effect that  the final scaling solution for the strings is increased beyond what would naively have been expected. Moreover, it also means that solutions that were previously found not to scale, now do enter scaling regimes as seen in Figs.\,(\ref{rhofigs_v}) and (\ref{figs_v_diffPs}). The consequences of this for observations are important. An increased scaling density by a factor of 10 or so as seen in our analysis implies that there is more string than previously assumed. It is important that the consequences of this are followed up.  We are currently determining the quantitative effects of these constraints on CMB anisotropies~\cite{inprep}.

\section*{Acknowledgements}
AA was supported by the Spanish National Research Council (CSIC),  FP7-PEOPLE-2007-4-3 IRG and a CTC Fellowship.  EJC is grateful for 
support from the Royal Society through a Wolfson Merit Award.  We would 
like to thank the Galileo Galilei Institute for Theoretical Physics and the Benasque Science Centre Pedro Pascual, where this work was initiated, for their warm hospitality.  EJC would like to thank Jaume Garriga and Roberto Emparan for their hospitality during his visit to the University of Barcelona.

\end{document}